\documentclass[9pt,twocolumn,twoside]{osajnl}
\journal{optica}
\setboolean{shortarticle}{true}
\usepackage{graphicx}
\usepackage{dcolumn}
\usepackage{bm}
\usepackage{soul,color}
\usepackage{float}
\usepackage{multirow}
\usepackage{tabularx}
\usepackage{dblfloatfix}
\usepackage{makecell}
\usepackage[flushleft]{threeparttable}

\title{Efficient and tunable blue light generation using lithium niobate nonlinear photonics}

\usepackage{upgreek}

\newcommand{\CMT}[1]{{}}

\author{Ayed Al Sayem}
\author{Yubo Wang}
\author{Juanjuan Lu}
\author{Xianwen Liu}
\author{Alexander W. Bruch}
\author{Hong X. Tang*}
\affil[1]{Department of Electrical Engineering, Yale University, New Haven, CT 06511, USA}
\affil[*]{Corresponding author: hong.tang@yale.edu}

\begin{abstract}
Thin-film lithium niobate (LN) has recently emerged as a playground for chip-scale nonlinear optics and leads to highly efficient frequency conversions from near-infrared to near-visible bands. For many nonlinear and quantum photonics applications, it is desirable to operate deep into the visible band within LN's transparency window. However, the strong material dispersion at short wavelengths makes phase-matching difficult, necessitating sub-micron scale control of domain structures for efficient quasi-phase-matching (QPM). Here we report the operation of thin film LN in the blue wavelength and high fidelity poling of thin-film LN waveguide to this regime. As a result, quasi-phase matching is realized between IR (871nm) and blue (435.5nm) wavelengths in a straight waveguide and prompts strong blue light generation with a conversion efficiency $2900\pm400\%$/W/cm$^2$. 
\end{abstract}

\setboolean{displaycopyright}{true}

\begin{document}

\maketitle

 \section{Introduction}
Integrated visible photonics has become a vastly growing area in optics \cite{blumenthal2020photonic} due to emerging applications such as atomic clocks \cite{newman2019architecture}, quantum information processing with quantum dots \cite{hepp2019semiconductor}, trapped ions and atom-like color centers \cite{sorace2018multi, mehta2020integrated, greentree2008diamond},  bio-sensing \cite{wilson2005advances} and so on. 
Among many integrated photonic material platforms \cite{SiNhosseini2009high,Morin:21, SiO2lee2017towards, liu2018ultra, west2019low}, thin-film $\mathrm{LiNbO_{3}}$-on-insulator (LNOI) stands out for its ultra-low optical loss \cite{ring_loncar, loncardesiatov2019ultra}, favorable electro-optic effect \cite{mod_loncar1} and strong optical $\chi^{(2)}$ nonlinearity \cite{poling_loncar,lu2020toward}. Recently, short wavelength operation on thin-film LN platform has been demonstrated at 633\,nm \cite{loncardesiatov2019ultra}. Other than low loss and high optical nonlinearity, a particular attraction of LN is the ability to reverse the domains periodically for achieving quasi phase matching (QPM) through electric field induced poling \cite{poling_loncar,fejerjankowski2020ultrabroadband,lu2020toward}. Such periodically-poled thin-film LN waveguides provide further enhanced nonlinear coupling due to the high mode confinement and have been recently exploited for efficient photon pair generation \cite{zhao2020high}, quantum frequency conversion \cite{QFC_ICFO,chen2021photon}, second harmonic generation in near-visible wavelengths \cite{loncardesiatov2019ultra,Lu2019,lu2020toward,Chen2019}. Despite these significant advances, deep visible second harmonic (SH) generation especially at the blue wavelength has remained elusive due to the stringent requirement for ultra short poling period with sub-micron domain structures \cite{Subzhao2020poling} and meanwhile maintaining low propagation loss at these wavelengths. 


Here we demonstrate  high fidelity control of ferroelectric domain structures in a z-cut LN nano-photonic waveguide, which enables first order quasi-phase matching for near IR to blue second harmonic generation with a poling period ($\sim2\,\upmu\mathrm{m}$) and 40\% duty cycle. By exploiting the strongest nonlinear co-efficient $\mathrm{d_{33}}$, we show bright, efficient blue light generation with a conversion efficiency consistent with theoretical limit. Our advance of operating thin film LN to the lower wavelength edge of its transparency window makes this nonlinear material platform particularly attractive for visible photonic applications in classical and quantum regimes. To highlight this utility, we generate blue SH at 435.5\,nm which targets the transition wavelength for Ytterbium-ion clocks with an on-chip efficiency 2900$\pm$400\%/W/cm$^2$. To the best of our knowledge, this is a record efficiency for blue second-harmonic generation in an integrated thin film material platform. 
Besides SH generation, we also assess the linear loss of LN in this new operating wavelength regime, which is found to be $3.9\pm0.5\,$dB/cm. Unlike resonant PPLN doublers which display strong photo-refractive effect \cite{Lu2019,lu2020toward,surya2021stable,jiang2017fast}, we show stable device operation even in deep visible band. Temperature control of the waveguides further allows the tuning of phase matching window over 3\,nm wavelength span (6\,nm at fundamental wavelength) with a 0.068\,$\mathrm{nm/^oC}$ tuning rate, ensuring their practical applications that require precision alignment with specific atomic transitions.

\section{Device design and fabrication}

\begin{figure*}
 \centering\includegraphics[width=15cm]{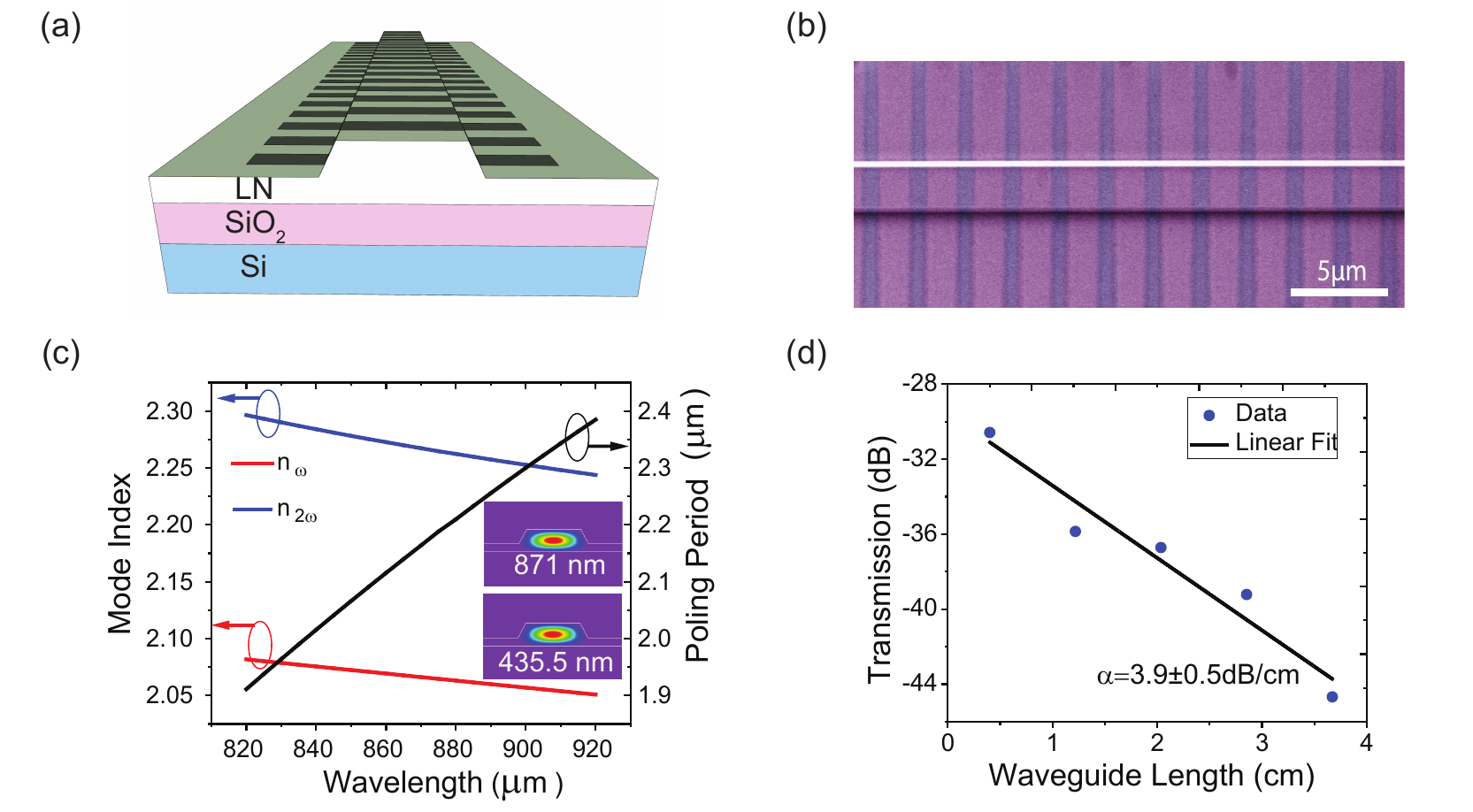}
\caption{(a) Schematic of the periodically poled thin-film Z-cut LN waveguide. (b) False-color scanning electron microscopy (SEM) image of a section of the poled waveguide. The poled region is marked with darker color. (c) Simulated effective mode index and poling period as a function of wavelength with a fixed film thickness of 605\,nm and etch depth of 375\,nm. The inset shows the Simulated electric field distribution for the fundamental TM mode (top) and second harmonic TM mode (bottom).(d) Transmission measured as a function the length of the LN waveguide at $\sim$435.5\,nm, with linear fitting (solid curve) showing an estimated $3.9\pm0.5\,$dB/cm propagation loss. 
} 
\label{Fig-Simulation}
\end{figure*}

Fig.\ref{Fig-Simulation}(a) shows the schematic of the PPLN device. The top width of the fabricated ridge waveguide was $1.8\,\upmu\mathrm{m}$ with an etching depth of 375\,$\mathrm{nm}$ and the length of the poled section was 6\,$\mathrm{mm}$. The poled device is imaged with a scanning electron microscope which directly reveals the domain structures (Fig.\ref{Fig-Simulation}b). Fig.\ref{Fig-Simulation}(c) plots the simulated effective mode index for the fundamental and second harmonic modes along with the required poling period. The poling period is varied around $2\,\upmu\mathrm{m}$ which is among the shortest poling periods achieved with thin-film LN. \cite{poling_optoca_2016,poling_loncar,zhao2020high, Lu2019,mishra2021midinfrared,Jankowski2020,zhao2020shallow}. 

\begin{figure*} [hbt!]
   \centering\includegraphics[width=18cm]{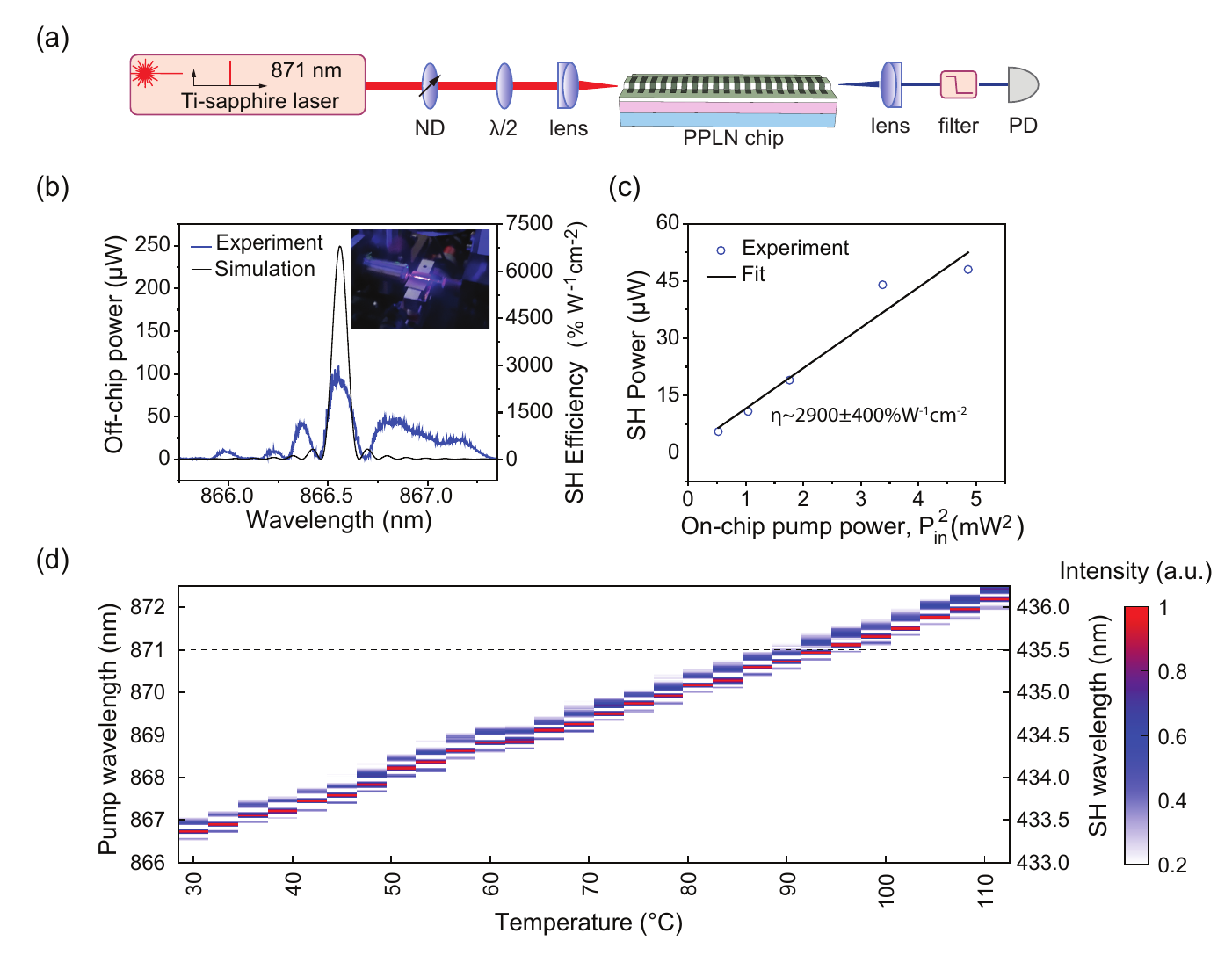}
\caption{(a) Schematic measurement setup. (b) SH output power and corresponding on-chip SHG efficiency along with simulation values as the fundamental wavelength is varied. The inset shows an image of the generated blue SH light during measurement. (c) SH output power as a function of the input pump power. (d) Temperature tuning of the peak SH wavelength crossing the $^2\mathrm{S}_{1/2}$ to $^2\mathrm{D}_{3/2}$ transition wavelength of $^{171}Yb^+$ ion clock.
}
\label{Fig-Results}
\end{figure*}

The device is patterned from a commercial thin-film LN on insulator (LNOI) wafer (supplied by NANOLN) with 605\,$\mathrm{nm}$-thick MgO doped $Z$-cut LN film bonded to $2\,\upmu\mathrm{m}$ thick silicon dioxide ($\mathrm{SiO_{2}}$) on a $400\,\upmu\mathrm{m}$-thick silicon handle. The fabrication process of the LN waveguide closely follows our previous work \cite{Lu:21,sayem2020lithium}. To achieve high fidelity poling, a thin layer of $\mathrm{HfO_{2}}$ ($\sim$\,10\,nm) is deposited on top of the fabricated photonic device using the atomic layer deposition (ALD) technique, with which the thickness of $\mathrm{HfO_{2}}$ can be controlled with nano-meter precision and keeps the poling parameters invariant in different fabrication runs. Since $\mathrm{HfO_{2}}$ is a high-k material, it also aids in confining the electric field during poling which is necessary for ensuring a good duty cycle. 
After oxide deposition, Ni electrodes were deposited and patterned on top of the waveguide using commonly used PMMA and lift-off process. 
An optimized poling sequence was then applied to create the desired poling pattern\cite{Lu:21,lu2020toward}. Afterwards the Ni electrodes and the interface oxide were sequentially removed by wet etching. Figure \ref{Fig-Simulation}(b) shows the false-color SEM image of the poled waveguide, where an excellent poling fidelity with reversed domains as small as 0.8\,$\upmu\mathrm{m}$ and a duty cycle close to $40\%$ are indicated. 

\section{Experimental results}

Before evaluating the nonlinear conversion efficiency, we first investigate the linear propagation loss which is a crucial figure of merit for thin-film integrated LN nano-photonic devices at short wavelengths. 
We assess the linear propagation loss of thin-film LN waveguides at 435.5\,nm using the cut back approach \cite{Morin:21}. Meander waveguides with the same waveguide cross-section as the PPLN nonlinear devices of varying lengths were fabricated and two identical lensed fibers were employed to measure the transmission. Fig. \ref{Fig-Simulation}(d) shows the measured transmission as a function of the length of the waveguide. The linear fitting provides an estimated value of the propagation loss to be $3.9\pm0.5\,$dB/cm, which is comparable to the reported value of another nonlinear thin-film material, AlN \cite{liu2018ultra}. 

The experimental setup used for SHG measurement is illustrated in Fig. \ref{Fig-Results}(a). A Ti-sapphire laser (M2 SolsTiS, 700–1000\,nm) is used as a tunable source of input light near $\sim$871\,nm, whose wavelength is precisely calibrated with a high resolution (0.1\,pm) wavemeter. A variable neutral density filter (ND) is employed to adjust the input power. The incident light is controlled to be vertically polarized with a half-wave plate and then launched into the waveguide using an aspheric lens. The output light from the chip is collected using another free space lens, from which the pump is filtered via an edge pass filter and then sent into a visible photo-detector (PD). To calibrate the efficiency, the free space filter is removed in order to measure the total transmission of the input light at $\sim$871\,nm.

Figure \ref{Fig-Results}(b) shows the measured SH power and the normalized SH efficiency as a function of the pump  wavelength. Based on the calibrated insertion loss of 15\,dB/facet, the experimental peak efficiency reaches 2900$\pm$400\%/W/cm$^2$ which is lower than the theoretical efficiency by $2.5\,$ times as indicated in the black curve. The most likely reason for the lower than the theoretical efficiency and also the undesired side-lobes is the inhomogeneity of the film thickness, which falls within the range of a few nm \cite{poling_optoca_2016,poling_loncar,zhao2020high}. The thickness variation causes the phase-matched poling period to vary along the length of the PPLN waveguide thus broadening the output spectrum. Experimentally, we observed the SH peak to vary $\sim0.21\,$nm per nm change of the poling period. From simulation we estimate the poling period to vary $\sim2.25\,$nm per nm variation of the total thickness. So 1\,nm variation of the film thickness along the PPLN waveguide can shift the SHG spectrum by $\sim0.5\,$nm thus drastically effecting the SHG spectrum and the SHG efficiency. The impact of film thickness variation is more prominent at deep-visible short wavelengths because of strong dispersion in this regime, which leads to large variations of poling period. The domain size variation along the waveguide due to poling non-uniformity may also influence the SH conversion efficiency due to the much shorter poling period utilized here.

 The inset of Fig.\ref{Fig-Results}(b) shows the image of bright blue light generated on-chip during the measurement. Figure \ref{Fig-Results}(c) shows the measured off-chip SH power as a function of on-chip pump power. 


The maximum output power of current devices is limited by the insertion loss, which prevents us from sending more pump power to the device without damaging the facet \cite{ledezma2021intense}. The high insertion loss in visible photonic waveguides is mainly due to the modal area mismatch between the coupling lens and the waveguide facet. Insertion loss could be significantly reduced by utilizing techniques typically applied in the near-infrared wavelengths such as inverse tapering with the fully etched waveguides or double etched waveguides incorporating mode converters \cite{hu2021high,he2019low} and will be implemented in our future work. 

At last, we show that our SHG device can target specific atomic transitions such as Ytterbium ion (Yb$^{+}$) clock's  $^2\mathrm{S}_{1/2}$ to $^2\mathrm{D}_{3/2}$ transition at 435.5\,nm \cite{hinkley2013atomic}. By controlling the temperature of the waveguide chip, we can fine adjust the average phase matching condition and tune the SH wavelength to this clock transition. Figure\,\ref{Fig-Results}(d) presents the precise SH wavelength tuning of our PPLN device from 867\,nm to 872\,nm when varying the chip temperature at a rate of 0.068\,$\mathrm{nm/^oC}$. By combining a miniaturized 871\,nm semiconductor laser of Hz-level linewidth available today \cite{Lai2021}, an efficient PPLN SHG device, and an on-chip heater for temperature tuning, ultra narrow-linewidth optical local oscillator operating at 435.5\,nm can be constructed for future integrated Yb+ ion clock applications.  
 
\section{Conclusion}

In conclusion, we demonstrate an efficient on-chip blue second-harmonic generator in thin-film integrated platform using high-fidelity first order PPLN waveguides. Low propagation loss at the deep-visible wavelengths has also been demonstrated, which may open up the possibility of integrating many passive, nonlinear and electro-optic components on the same chip for critical applications such as compact atomic clocks. At the same time, the nonlinear thin-film LN platform could find promising classical and quantum photonic applications such Lidar, virtual reality, quantum memory, long range quantum communication, and quantum computing devices based on trapped ions and atoms.

\vspace{2 mm}
\begin{backmatter}
\bmsection{Funding} This project is supported by Sandia through DARPA's A-PHi program under proposal 052181213-1 (DENA00003525). We acknowledge partial funding from DOE (DE-SC0019406) for materials used in this experiment. HXT acknowledges support from  National Science Foundation (EFMA-1640959).

\bmsection{Acknowledgments}
We thank Drs. Yu-Hung Lai, Paul Parazzoli, and Hayden McGuinness for valuable discussions. The facilities used for device fabrication were supported by the Yale SEAS Cleanroom and the Yale Institute for Nanoscience and Quantum Engineering (YINQE). The authors would like to thank Dr. Yong Sun, Dr. Michael Rooks, Sean Rinehart, and Kelly Woods for their assistance provided in device fabrication.

\bmsection{Disclosures} The authors declare no conflicts of interest.

\bmsection{Data availability} The data that  support the findings of this study are available from the corresponding authors upon reasonable request.

\end{backmatter}


\bibliography{sample}

\bibliographyfullrefs{sample}

\end{document}